\documentstyle[12pt,aaspp4]{article} 
\def\be{\begin{enumerate}}
\def\ee{\end{enumerate}}
\def\bi{\begin{itemize}}
\def\ei{\end{itemize}}
\def\bd{\begin{description}}
\def\ed{\end{description}}

\def\etal{{\it et al.~}}
\def\myarcmin{^\prime\mskip-5mu}
\def\EO{{\it Einstein Observatory}}
\def\einstein{{\it Einstein}}
\def\ginga{{\it Ginga}}
\def\rosat{{\it ROSAT}}
\def\exosat{{\it EXOSAT}}

\def\asca{{\it ASCA}}

\def\la{\hbox{\raise.35ex\rlap{$<$}\lower.6ex\hbox{$\sim$}\ }}
\def\ga{\hbox{\raise.35ex\rlap{$>$}\lower.6ex\hbox{$\sim$}\ }}

\received{12 September 1996}
\accepted{29 May 1997}

\slugcomment{Accepted for publication in {\it The Astrophysical Journal}}

\lefthead{Harrus et al.}
\righthead{Interpretation of W44}

\begin{document}
\title{Interpretation of the Center-Filled Emission from the
Supernova Remnant W44}

\author{Ilana M.~Harrus\altaffilmark{1,2},  
John P.~Hughes\altaffilmark{1,3},
K.~P.~Singh\altaffilmark{4},
K.~Koyama\altaffilmark{5},
and I.~Asaoka\altaffilmark{6}
\altaffiltext{1}{Harvard-Smithsonian Center for Astrophysics, 
                 60 Garden Street,  Cambridge MA 02138} 
\altaffiltext{2}{Department of Physics, Columbia University, 
                 550 W 120$^{\rm th}$ Street, New York, NY 10027}
\altaffiltext{3}{Department of Physics and Astronomy, Rutgers
                 University, P.O.~Box 849, Piscataway, NJ 08855-0849}
\altaffiltext{4}{Tata Institute of Fundamental Research, 
                 Homi Bhabha Road, Mumbai, 400 005 India}
\altaffiltext{5}{Department of Physics, Kyoto University, 
                 Kitashirakawa, Oiwake-cho, Sakyo-ku, Kyoto 606, Japan}
\altaffiltext{6}{Max-Planck-Institute f\"ur Extraterrestrische Physik, 
                 W-8046 Garching  bei M\"unchen, Germany}}

\begin{abstract}

\par\indent

We report the results of spectral and morphological studies of X-ray
data from the supernova remnant (SNR) W44.  Spectral analysis of
archival data from the \EO, \rosat, and \ginga, covering a total
energy range from 0.3 to 8~keV, indicates that the SNR can be
described well by a nonequilibrium ionization (NEI) model with
temperature $\sim$0.9 keV and ionization timescale of order $6000$
cm$^{-3}$ years. All elemental abundances are found to be within 
about a factor of two of their cosmic values, with iron possibly appearing
to show significant depletion. No clear evidence for emission from
supernova ejecta can be inferred from the observed metal abundances.
The column density toward the SNR is high -- around~10$^{22}$ atoms
cm$^{-2}$ -- as expected given the location of the remnant in the
Galactic plane.

In addition to the spectral analysis, we have investigated two
different evolutionary scenarios to explain the centrally-brightened
X-ray morphology of the remnant: (1) a model involving the slow
thermal evaporation of clouds engulfed by the supernova blast wave as
it propagates though a clumpy interstellar medium (ISM) (White \& Long
1991, hereafter WL), and (2) a hydrodynamical simulation of a blast
wave propagating through a homogeneous ISM, including the effects of
radiative cooling.  Both models can have their respective parameters
tuned to reproduce approximately the morphology of the SNR. The mean
temperature of the hot plasma in W44 as determined by our NEI X-ray
analysis provides the essential key to discriminate between these
scenarios.  Based on the size (using the well established distance of
3 kpc) and 
temperature of W44, the dynamical evolution predicted by the WL
model gives an age for the SNR of merely 6500 yr.  We argue
that, because this age is inconsistent with the characteristic age
($P/2\dot P \sim 20000$ yr) of PSR 1853+01, the radio pulsar believed
to be associated with W44 (Wolszczan, Cordes, \& Dewey 1991),
\nocite{WOLSZ91} this model does not provide the explanation for the
center-filled morphology.  We favor the radiative-phase shock model since
it can reproduce both the morphology and age of W44 assuming reasonable
values for the initial explosion energy in the range $(0.7 - 0.9)
\times 10^{51}$~ergs and the ambient ISM density of between 3 and
4 cm$^{-3}$.

\end{abstract}

\keywords{ISM: abundances -- ISM: individual (W44) -- nuclear
reactions, nucleosynthesis, abundances -- shock waves -- supernova
remnants -- X-rays: ISM}

\newpage

\section{ Introduction }
\par\indent

The study of supernova remnants (SNRs) provides a unique tool with
which to deepen our understanding of the interstellar medium (ISM) and
the processes which shape its structure, energetics, and composition.
At the time of their death, stars induce a formidable release of
energy into the ISM and a strong shock wave begins to propagate.  At
early stages of the evolution, the observed emission contains
contributions from both the supernova (SN) ejecta and the ISM.  One of
the major challenges for X-ray spectroscopy of young SNRs is the
differentiation and characterization of these two contributions.
Knowledge of the composition of SN ejecta is of considerable
importance for the constraints it can provide, through the study of
nucleosynthesis, on the nature and evolution of the progenitor star.
At later stages, SNRs and their evolution are dominated by the
interstellar medium. In this article we explore the nature of W44,
a middle-aged supernova remnant which is in the ISM-dominated stage of
evolution. 

W44 was first discovered as a radio source in a survey by Westerhout
(1958), \nocite{WESTER58} and was observed later by several others
(Mills, Slee, \& Hill 1958; Edge et al.~1959).
\nocite{MILLS58} \nocite{EDGE59} It was identified as a possible
supernova remnant by Scheuer (1963) \nocite{SCHEU63} because of its
shell-like radio structure and its non-thermal radio spectrum.
OH and H {\sc i} absorption measurements (Goss 1968;
Goss, Caswell, \& Robinson 1971; Radhakrishnan et al.~1972; and Knapp
\& Kerr 1974) \nocite{GOSS68} \nocite{GOSS71} \nocite{KNAPP74}
\nocite{RADH72} have resulted in a more complete mapping of the
heavily obscured surroundings of the SNR and lead to the widely
accepted distance to the remnant of 3 kpc.  The 20~cm VLA image
shows a roughly elliptical limb-brightened radio shell with major and
 minor semi-axes of 15 pc and 10 pc for this assumed distance
(Jones, Smith, \& Angellini 1993). \nocite{JON93} Knots and filaments
contributing to the emission are also seen; Jones et al.\ (1993) have
interpreted these as arising from the radiative shocks driven into
interstellar clouds.  The radio emission is non-thermal with a
spectral index of $-$0.3 and it is highly polarized ($>$ 20\%; Kundu
\& Velusamy 1972)
\nocite{KUND72}.

There are two radio pulsars in the vicinity of W44.  One of them, PSR
1854+00 (Mohanty 1983), \nocite{MOHEN83} is old ($10^8$ yr), which
makes an association between it and the remnant unlikely.  The
discovery of the other pulsar, PSR 1853+01, is more recent (Wolszczan,
Cordes, \& Dewey 1991). \nocite{WOLSZ91} Taylor, Manchester, \& Lynn
(1993) \nocite{TAYL93b} give a distance to this pulsar of
3.3$\pm$0.3~kpc, while the dispersion-measure distance (Taylor \&
Cordes 1993) \nocite{TAYL93b} is 2.8$\pm$0.1~kpc, both of which are in
excellent agreement with the distance estimate to W44 mentioned above.
There is a faint radio synchrotron nebula associated with PSR 1853+01 (Jones
et al.~1993; Frail et al. 1996); an X-ray counterpart to
the nebula has also been announced recently (Harrus, Hughes, \& Helfand
1996) although it accounts for only 3.3\% of the total X-ray luminosity of the 
remnant in the 0.4--2.0~keV band.  
This is a young pulsar: it exhibits large amplitude timing noise, and the
ratio of the period to period derivative, the characteristic spin-down
age of the pulsar, is $\sim$20000 years.  Assuming that one can reliably
associate the pulsar and the SNR W44, this provides an independent
estimate of the remnant's age, which, as we show below, offers an
extremely valuable piece of information for discriminating between
evolutionary scenarios.

W44 was discovered as an X-ray source by the Astronomical Netherlands
Satellite (Gronenschild et al.~1978), and has been a popular target of
all subsequent X-ray astronomy satellites. Smith et al.\ (1985)
\nocite{SMIT85} presented the first detailed X-ray imaging
observations based on data from the imaging proportional counter (IPC)
onboard the {\it Einstein Observatory}.  In the soft X-ray band, W44
presents a centrally-peaked morphology, reminiscent of a pulsar-driven
synchrotron nebula (like the Crab Nebula) than a shell-type SNR (like
the Cygnus Loop). However, the X-ray spectrum of W44 (Jones et
al.~1993; Rho et al.~1994) \nocite{RHO94} is predominantly thermal in
origin, based on the presence of strong emission lines from highly
ionized atoms of magnesium, silicon, sulfur, and iron clearly observed
by the \einstein\ Solid State Spectrometer (SSS).  Notwithstanding the
presence of PSR 1853+01 and its associated synchrotron nebula, W44
belongs to the class of ``center-filled'' remnants which show
limb-brightened radio shells, centrally-peaked X-ray morphologies, and
predominantly thermal X-ray spectra.

In the Sedov (1959) model of supernova remnant evolution, the SN blast
wave propagates through an isotropic homogeneous ISM.  At the
shock front, the swept-up material is heated to temperatures of order
10$^7$ K and results in a shell-like X-ray morphology with a thermal
spectrum. The limb-brightened radio emission comes from compression of
the interstellar magnetic field and the accompanying acceleration of
electrons which also occurs at the SNR shock front.  This simple
model, although apparently successful for many of the known
shell-like SNRs, fails to account for remnants such as W44, which have
a distinct, centrally-peaked X-ray morphology.

Other models have been proposed to explain the observed morphology of
W44 and other remnants of this type. In one particular scenario, the
remnant is in a later phase of evolution when the blast wave has gone
radiative (shock velocities of roughly 300 km s$^{-1}$; Cox 1972). The
radio shell traces the position of the shock front, but, since the
X-ray emission from the shock front is soft ($kT\sim 10^6$ K) and the
line-of-sight ISM column density is significant ($\sim$10$^{22}$ atoms
cm$^{-2}$), the outer shell is essentially invisible in the X-ray band
(Smith et al.\ 1985).  The X-ray emission comes rather from the hot
interior of the remnant, providing a center-filled morphology.  This
model predicts that W44 should be old and that the X-ray temperature
should decrease from the center of the SNR toward the edge.  Some
support for this scenario comes from the recent discovery of H$\alpha$
and S {\sc ii} emitting optical filaments around the periphery of the
X-ray emission (Rho et al.~1994) and an expanding shell of H {\sc i}
emission (Koo \& Heiles 1995), indicating the presence of cool gas
there. In the White \& Long (1991) (WL) scenario, the SNR is expanding
into a cloudy ISM (as in the ISM models of McKee
\& Ostriker 1977) and evaporating clouds produce an increased density
of hot gas in the interior, giving the centrally-peaked appearance.
Here the temperature is expected to be relatively uniform throughout
the interior. 

In this article we explore the implications of these two evolutionary
scenarios for W44 using constraints obtained from X-ray imaging and
spectroscopic observations from the {\it Einstein Observatory}, {\it
ROSAT}, and {\it Ginga}.  The next section consists of a brief summary
of our current knowledge of the SNR in the X-ray band, and of the
observations used in our analysis.  Our models and the analysis
techniques which provide the relevant observational constraints are
presented in \S~3.  We apply these constraints to W44 in the context
of the two proposed evolutionary scenarios in \S~4.  A summary of the
paper's main points is to be found in \S~5.

\section{X-ray Observations of W44}
\subsection{\rosat\ PSPC}

The first set of data to be described comes from the position
sensitive proportional counter (PSPC) (Pfeffermann et al.~1986)
onboard \rosat\ (Tr\"{u}mper 1983). \nocite{TRUEMP83}
\nocite{PFEFF86} The PSPC spectral resolution ($\Delta E/E$) was about
45\% (FWHM) at 1 keV and the instrument was sensitive over the energy
band 0.1--2.4 keV.  W44 was observed by the PSPC in April 1991.  Note
that these data were also analyzed by Rho et al.\ (1994).

We extracted the PSPC data from the \rosat\ archive and carried out
the following reduction procedures.  First we applied a time filter to
reject data during orbital periods contaminated by solar X-rays
scattered into the telescope field of view by the upper
atmosphere. These periods manifest themselves as sudden increases in
the total count rate at the beginning or end of the good-time
intervals supplied as part of the \rosat\ standard processing.  After
rejecting these time periods, the deadtime-corrected exposure time was
6726 s.  In order to minimize contamination due to particle-induced
background, it has been recommended that data be rejected during time
intervals when the master veto rate is greater than 170~s$^{-1}$
(Snowden et al.~1994). \nocite{SNOW93} Since only 3\% of our remaining data
had an associated master veto rate above this threshold, we decided
not to apply an additional time filter to reject those events.  We
also restricted our analysis to the central 40$^\prime$ region of the
detector.

The W44 source spectrum was extracted from within the region defined
by the surface brightness contour corresponding to 10\% of the peak
brightness.  The background region lay outside this, but still came
from within the central region of the PSPC (within the window support
ring).  The background spectrum was corrected for the energy-dependent
difference in detector response (mainly due to off-axis vignetting)
between the source and background regions and was normalized by the
ratio of solid angle between the regions.  After background
subtraction, the total PSPC count rate of W44 is $4.22 \pm 0.02$
s$^{-1}$.  The PSPC spectrum of the entire remnant is shown in Figure~1.
A fit to these data using a solar-abundance, collisional equilibrium
ionization thermal plasma model (Raymond \& Smith 1977; 1992 July 27
version, hereafter RS) provided an unacceptable fit with a $\chi^2$ of
39 for 19 degrees of freedom.  In this case, the best-fit
temperature, $kT$, was $\sim$1 keV and the column density, $N_{\rm
H}$, was $7.9 \times10^{21}$ cm$^{-2}$.

\subsection{\einstein\ SSS}

The \einstein\ solid-state spectrometer (SSS) has been described in
detail by Joyce et al.\ (1978) and Giacconi et al.\ (1979);
\nocite{JOYC78} \nocite{GIAC79} here we provide only a brief
discussion of its main characteristics. The SSS was sensitive to
X-rays between 0.4--4.0 keV with a nominal spectral resolution (FWHM)
varying from 30\% at low energies to 4\% at high energies.  During
orbital operations, an unexpected problem of ice formation on the
detector window occurred, which caused a time dependence in the low-energy 
efficiency of the SSS. An empirical model for this effect has
been developed based on the analysis of a number of observations of
the Crab Nebula taken throughout the course of the \einstein\ mission
(Christian et al.\ 1992); \nocite{CHRIS92} in this work we employ the
nominal ice absorption model appropriate to the dates of observation
of W44.

The total SSS exposure time on W44 was 22,608~sec. The data were
acquired in four separate pointings toward two different regions of
the remnant.  (Note that the field of view of the SSS was about
$6^\prime$ in diameter and thus a single pointing did not cover the
entire SNR.) The four datasets were compared and, since they were
consistent with each other within the statistical errors, they were
summed to form a single spectrum. The separate response functions were
averaged (weighting by each pointing's exposure time). A total of 8072
source photons were detected. In order to account for systematic
uncertainties in the ice absorption model, as well as other
uncertainties in the SSS calibration, we have added a systematic error
equal to 2\% of the source intensity in each spectral bin.  The
minimum energy we consider for this data set is 0.8~keV. The SSS
spectrum in Figure~1 shows obvious emission lines from K$\alpha$
transitions of highly ionized atoms of magnesium, silicon, and sulfur,
which clearly points to a thermal origin for the X-ray emission.
Nevertheless this spectrum cannot be fitted well by a simple solar
abundance RS thermal plasma emission model.  The reduced $\chi^2$ of
4.5 obtained in this case for $kT \sim 0.9$ keV and $N_{\rm H} =
7.4\times 10^{21}$ cm$^{-2}$ indicates that a more detailed analysis,
including effects such as nonequilibrium ionization, is necessary.

\subsection{\ginga\ LAC}

The major experiment on \ginga\ (Makino et al., 1987) \nocite{MAKI87}
was the Large Area Counter (LAC) (Turner et al. 1989), \nocite{TURN89}
an array of eight sealed proportional counters with a total geometric
collecting area of 4000 cm$^2$, mechanically collimated to a field of
view of roughly 1$^\circ$ by 2$^\circ$ (FWHM).  The efficiency of the LAC for
collecting X-rays was greater than 10\% over the 1.5--30 keV
band. The lower-energy limit was defined by the thickness of the Be
window material ($\sim$62 $\mu$m), while the high-energy limit arose
from the finite active depth of the proportional-counter gas
volume. These detectors had an energy resolution of 18\% (FWHM) at
about 6 keV. With its very low internal background rate and large
effective area, the LAC was a very sensitive instrument for carrying
out X-ray spectral studies.

No direct pointing of W44 was made by \ginga. Rather we have extracted
data on the source from a scan of the Galactic plane carried out on 12
September 1988. Scan data were taken in MPC2 mode, which combined the
data from the top and middle layers of the LAC and summed the data
from four detectors into one before telemetering to the ground. The
two spectra so obtained were summed during data reduction. Background
was determined from source-free regions of the scan on either side of
W44.  The effective exposure time was low (1984 s); the source
is bright however, ($\sim$15 counts s$^{-1}$), and the X-ray spectrum is
well-defined from 1.5 keV to 10 keV.  The spectrum is soft, consistent
with a RS thermal model with $kT \sim 0.75$ keV and an interstellar
column density of $\sim$10$^{22}$ cm$^{-2}$.  There is no evidence for
any harder emission component in the \ginga\ data.  We set an upper
limit (3 $\sigma$) of $3.6 \times 10^{-12}$ ergs cm$^{-2}$ s$^{-1}$ to
the 2--10 keV flux of a Crab-like power-law component ($dN/dE \sim
E^{-2.1}$) contributing to the \ginga\ spectrum of W44.

\subsection{Other Observations}

During the initial phases of this study, we explored the possibility
of using X-ray observations of W44 from other sources of archival
data, namely from the \EO\ and \exosat. After careful evaluation it
became clear that these data would not be useful in our study. We
review our arguments for arriving at this conclusion below.

The \einstein\ imaging proportional counter (IPC) is similar to the
\rosat\ PSPC in many respects.  The major advantage of the IPC over
the PSPC is its higher-energy cutoff (4.5 keV vs.\ 2.4 keV).  This
advantage, however, is largely offset by the IPC's poorer energy
resolution and image quality, and the large uncertainty in its
calibration which limits its usefulness for detailed spectral
analysis. In our preliminary spectral fits, it was found that the IPC
global spectrum was consistent with that from the PSPC and SSS,
although the best-fit $\chi^2$ for the IPC data was formally
unacceptable. Although consistent with the other data, the IPC
spectrum does not provide additional constraints on the model and thus
we reject it as being redundant.

Data from the medium energy (ME) proportional counters on \exosat\ are
available through the High Energy Astrophysics Science Archive
maintained by the Goddard Space Flight Center (GSFC).  The data in
this archive have undergone a standard reduction procedure to produce
background-subtracted spectral files for analysis.  The processing
flag for the ME spectrum of W44 is listed as quality 2, which
indicates a major problem with the reliability of the data.  Indeed the
ME spectrum of W44 shows a hard component above about 5 keV and a
reasonably strong K$\alpha$ iron line, both of which are entirely absent in
the \ginga\ LAC spectrum.

The problem with the standard background subtraction for the W44 data
was identified by Jones et al.\ (1993), who examined the raw ME
data and found that a significant fraction of it was contaminated by
irregular count-rate flares presumably induced by penetrating charged
particles.  After rejecting the data from the most seriously affected
detectors, Jones et al.\ (1993) obtained good fits to the ME data of a
single-temperature RS thermal model with $kT \sim 0.9$ keV and $N_{\rm
H} \sim 10^{22}$ cm$^{-2}$. These results are consistent with those
derived using the \ginga\ data.  Because we had no access to the raw
\exosat\ data and since the \ginga\ LAC covers the same energy band
and is therefore fully complementary, we decided to exclude the ME
data altogether.

In their complex analysis of W44, Rho et al.\ (1994) use the contaminated
ME data obtained directly from the GSFC archive. This explains why these
authors require a high temperature component in their model fits.
In our view, it also likely invalidates the conclusions they arrive at 
concerning
their best-fitting NEI spectral model (i.e., shock temperature,
ionization timescale, and assumptions about electron-ion temperature
equilibration timescales).

\section{Nonequilibrium Ionization Modeling and Analysis}

Accurate plasma diagnostics are the key to our understanding of the
physical phenomena which occur during supernova remnant evolution.  At
the simplest level, measurements of plasma temperature and elemental
abundances allow one to derive quantitative values for the plasma
density in the SNR from the intensity and brightness distribution
shown by broadband X-ray images.  Furthermore, as we show below, the
remnant's radius, temperature, and density are essential quantities
for understanding its dynamical state. The relative abundance ratios
of the X-ray emitting plasma, as determined by spectroscopy, can
indicate the presence of reverse-shocked ejecta, again providing clues
to the evolutionary state of the remnant. The driving force behind the
detailed spectral fits we pursue in the following section is the
derivation from the observational data of the most accurate values
possible for the thermodynamic quantities, of which the most
important is the mean electron temperature. 

Interpretation of SNR X-ray spectra is complicated by the
nonequilibrium processes that occur in low-density shock-heated
plasmas and which necessitate detailed time-dependent models of the
spectral emissivity. One important influence on the thermodynamic
state of the plasma is the fact that the ions are not instantaneously
ionized to their equilibrium configuration at the temperature of the
shock front. Rather, the timescale for attaining full equilibrium
ionization is comparable to the remnant dynamical timescale.  Numerous
authors have incorporated this nonequilibrium ionization (NEI) effect
into models of SNR spectral emissivity. Here we use the matrix
inversion method developed by Hughes \& Helfand (1985) to solve for
the time-dependent ionization fractions, and couple it to the RS
plasma emission code (see Hughes \& Singh 1994 for more details).  The
column density of neutral hydrogen along the line-of-sight, $N_{\rm
H}$, is included as a fit parameter using the cross sections and ISM
abundances from Morrison \& McCammon (1983). \nocite{MORR83}

\subsection{Single-temperature, single-timescale NEI model}

The simplest NEI model assumes that the X-ray emitting plasma was
impulsively heated to temperature $kT$ some time $t$ ago.  The
temperature, defined as the kinetic state of the electrons, is assumed
to remain constant. The ionization state depends on the product of
electron density and age: i.e., the ionization timescale, $\tau_i
\equiv n_et$. We refer to this as the single-temperature,
single-timescale NEI model and we apply it here to the spectra from
the entire remnant.

The observed spectra constrain the electron temperature in the SNR
mainly through the shape of the continuum emission, which, to first
order, is independent of the ionization state, equilibrium or
otherwise. (In fact a model independent analysis of these data, using
a parameterized bremsstrahlung function plus several gaussians to
describe the line emission, yields a similar value, $\sim$1 keV, for
the electron temperature.) The ionization timescale is determined by
the centroid energies of the various K$\alpha$ lines (which are, after
all, blends of lines from hydrogen-like ions, helium-like ions, etc.\
and so depend sensitively on the ionization state) that appear in the
spectral band: Ne, Mg, Si, and S, in particular. The relative
intensities of these emission lines derived from the model fits can,
in principle, constrain individual elemental abundances. In practice,
some of the individual contributions are not easily separated,
especially for those species that are primarily continuum
contributors.  Therefore, for the elements He, C, N, and O we have
fixed the abundance to the solar values relative to hydrogen.  The
abundances of the other elemental species were allowed to vary
freely. We adopt as solar abundances the values given by RS.

Figure~1 shows the data and best-fit NEI model obtained when the three
data sets are fitted jointly. The minimum $\chi^2$ is 137.4 for 92
degrees of freedom.  The $\chi^2$ associated with the PSPC data is
25.4 (22 data bins), with the SSS data is 87.0 (72 data bins), and
with the LAC data is 25 (11 data bins).  The overall normalization of
the spectral data provides a value for the emission measure of the hot
plasma in W44: $n_{\rm H}^2 V / (4\pi D^2) = (1.76 \pm 0.37 )\times
10^{13}$ cm$^{-5}$. We use a value of 1.09 for the ratio 
$n_e / n_{\rm H}$. The best-fit values for the global spectral
parameters are temperature, $kT = 0.88 \pm 0.14$ keV; ionization
timescale, $\tau_i = (2.0^{+4.3}_{-0.7})\times10^{11}$ cm$^{-3}$~s; and
column density, $N_{\rm H} = (1.0^{+0.6}_{-0.2})\times 10^{22}$
atoms~cm$^{-2}$.  The quoted error bars are at the 90\% confidence
level for three interesting parameters ($\Delta \chi^2 = 6.25$).
Figure~2 shows graphically how $\chi^2$ varies with each of these global
parameters (also allowing all other parameters to vary freely). Table
1 provides a numerical summary of the abundance results.  The first
column gives the best-fit elemental abundances, relative to their
cosmic values. The second column shows the errors in abundance
determined with the temperature, ionization timescale, and column
density fixed at their best-fit values. The remaining columns gives
the errors in abundance arising from the variation in the global
spectral parameters as shown in Figure~2.

The ionization timescale we derive for W44 is representative of that
from a middle-aged remnant (like N132D in the LMC, see Hwang et
al.~1993) and is indicative of a plasma that is underionized for its
temperature.  From the \rosat\ image, we estimate the mean electron
density in the hot plasma (see below) to be $\langle
n_e^2\rangle^{1/2} \simeq 0.4$ cm$^{-3}$.  Combined with the
ionization timescale, this suggests an ``age'' of order 15000 yr, in good
agreement with other estimates of the age of W44 and its 
associated pulsar PSR 1853+01 (see below).

\subsection{Radial Temperature Gradient}

Some evolutionary scenarios for SNRs (e.g., Sedov) predict significant
radial variations in the plasma temperature, while others (e.g., WL)
do not.  We have used the \rosat\ PSPC data to constrain the allowed
range of temperature variation in W44 in the following approximate
manner.  Two PSPC spectra were extracted: one from within a radius of
$6\myarcmin.7$ and the other outside this region.  Note that the boundary
between the regions was chosen so that each spectrum had roughly the
same total number of detected counts; 
the results are not sensitive to the exact
position of the boundary. Each PSPC spectrum was fitted to an
independent NEI model, and the sum of these NEI models was required to
fit the LAC and SSS data.  These latter two datasets were assumed to
be representative of the entire remnant (which is strictly correct for
the LAC data, but only approximately true for the SSS).  The NEI
models were constrained to have the same abundances, column density,
and ionization timescale with values fixed to the best-fit ones
determined above, while the temperatures and intensity normalizations
of the models describing the two regions were allowed to vary
independently.  We obtain a better fit if the inner region is somewhat
hotter than the outer region.  At 90\% confidence, the temperature of
the inner region is constrained to be between 10\% and 20\% higher
than the temperature of the outer region.  This analysis indicates
that there is little variation in temperature with position in the
remnant, a result that is consistent with previous studies (Rho et
al.~1994).

\subsection{Multiple NEI Components}

In addition to the simple single-temperature, single-timescale NEI
model discussed above, we also have investigated the possibility that
the X-ray emitting plasma in W44 is in a more complex state.  First,
we looked for evidence that the ionization state varies as a function
of elemental species.  In this study we had two ionization timescales
as free parameters: one for a particular individual species (Ne, Mg,
Si, S, Ar, and Fe each in turn) and the other for all
the remaining elemental species. Fits were carried out with the other
relevant spectral parameters ($kT$, $N_{\rm H}$, and abundances)
constrained to be the same for all species and allowed to vary freely.
The derived $\chi^2$ values were compared to the single-temperature,
single-timescale results to assess the significance of the
introduction of the new parameter.  None of the elemental species for
which we pursued this analysis showed a statistically significant
difference, suggesting that the various elements contributing to the X-ray
emission are uniformly mixed throughout the plasma.  

We also carried out fits of a two-component NEI model to the entire
spectrum to see whether our data require that the plasma in W44 be
multi-phase. The components had the same starting abundance as found in the 
one-component NEI analysis and identical absorbing
column density.  We assumed that the media were in pressure
equilibrium which allowed us to relate the ionization timescales and
temperatures as $\tau_{i,2} = \tau_{i,1}\, T_1 / T_2$. Note that we also
made the implicit assumption that the two components were shocked at
the same time.  With this condition, only two additional free
parameters were introduced: the second temperature and the ratio of
emission measures between the two media. We explored values for the
temperature of the second component from 0.5 to 5 keV and the ratio of
emission measure from 0.1 to 10. Over this range of parameter space,
no statistically significant reduction in $\chi^2$ was observed,
although equally good fits were obtained in many cases. Our data allow
a second component with $kT= 2$ keV only if its emission measure is
less than $\sim$3\% that of the main component.  The allowed emission
measure for the addition of a 5 keV component is even more restricted:
$<$0.5\% of the main component. Because of the significant
interstellar cutoff our limits on gas at temperatures with $kT < 0.5$
keV are rather weak.

\subsection{Volume, Density, Pressure, and Mass Estimates}

\par

In the soft X-ray band W44 is roughly elliptical in appearance with a
long dimension of $33^\prime$ and a short one of $20^\prime$.  We
estimate the volume of the remnant as an ellipsoid with principal axes
in the plane of the sky with sizes as observed. The length of the
third axis is some factor, $\alpha$, times the size of the observed
short dimension.  This corresponds to a volume, $V = 1.3\times 10^{59}
\alpha D_{\rm 3\, kpc}^{\rm 3}$ cm$^{-3}$, for the nominal distance to W44 of
3 kpc.

\par

The root-mean-square electron density can be determined simply from
the fitted emission measure (\S 3.1) and the volume.  We obtain

$$ \langle n_e^2\rangle^{1/2} = 0.42\, (\alpha f D_{\rm 3\, kpc})^{-1/2} \
  {\rm cm}^{-3}$$ 

\par\noindent
where $f$ is the volume filling factor of the hot plasma .
The uncertainty 
on $\langle n_e^2\rangle^{1/2}$ from errors in the fitted emission
measure alone is $\pm 0.06$ cm$^{-3}$.  The average thermal pressure in the
remnant is roughly $1.1\times 10^{-9}$ ergs cm$^{-3}$, assuming that
the ion and electron temperatures are equal.

\par
The mass of X-ray emitting hot plasma is given by

$$M = 56\, (\alpha f)^{1/2}\, D_{\rm 3\, kpc}^{5/2}\ M_\odot.$$

\subsection{Abundances}

Vancura et al.\ (1994) argue for the use of depleted abundances when
interpreting the X-ray spectra of SNRs, due to the long timescales for
grain destruction within the shock heated gas.  In their models, the
proportion of intact silicate and graphite grains remaining after
being engulfed by a blast wave depends strongly on the shocked column
$N_S$, but only weakly on the shock velocity. For W44, we approximate
$N_S$ by the product of the RMS density and the mean observed radius,
which gives a value $N_S\sim 1.5 \times 10^{19}$ cm$^{-2}$.  The
fraction of initially depleted mass remaining in the solid phase for
this shocked column is 30--45\% (Vancura et al.\ 1994), implying that
the observed abundances of Mg, Si, S, and Fe in W44 should be slightly
below solar (abundances of 60--80\%).  The relatively inert elements Ne and
Ar should show no depletion.

Our best fit to the X-ray spectrum of W44 implies elemental abundances
that are close to or somewhat below the solar values (except for
iron), and in general agreement with the picture sketched above. The
strongest apparent depletion is observed for iron, although we are
wary of this result due to uncertainties in the atomic physics of iron
L-shell emission.  It is also true that the iron abundance varies
dramatically with changes in the global spectral-fit parameters, as
clearly shown in Figure~2. For $N_{\rm H}$ values slightly higher than
the best fit one (but still within the allowed range), all derived
elemental abundances are within a factor of $\sim$2 of solar.

On the other hand it is slightly puzzling that the observed abundance
pattern shows no clear evidence for the presence of SN ejecta.  Models
of nucleosynthesis in massive stars (ZAMS masses of 13--25 $M_\odot$)
predict the ejection of, for example, 0.047--0.116 $M_\odot$ of Si and
0.026--0.040 $M_\odot$ of S (Thielemann et al.\ 1996).  In total for
W44 we observe only about 0.03 $M_\odot$ of Si and 0.01 $M_\odot$ of S
and most of this, we have just argued, can be attributed to the
swept-up interstellar medium. Perhaps this suggests that the
progenitor of W44 is less massive than 13 $M_\odot$, although a lower
bound of 8 $M_\odot$ seems necessary in order to produce the neutron
star of the associated PSR 1853+01 (Wheeler 1981). The
fate of the metals ejected by a SN is also complex: adiabatic cooling
during the initial free expansion phase, subsequent heating by the
reverse shock, radiative cooling, the disruption of the ejecta, and so
on.  As interesting as these issues are, addressing them is certainly
beyond the scope of this work and, observationally, will require data
of considerably higher spatial and spectral quality than available
now.

\section{Evolutionary State of W44}

\par

The radio image of W44 (Jones et al.\ 1993; Frail et al.\ 1996) shows
an elliptical-shaped limb-brightened morphology with a size of
$34\myarcmin.8$ $\times$ $24\myarcmin.4$. In the models we consider below
we use the boundaries of the radio emission to delineate the position
of the blast wave, employing the mean radius $R_s = 13.1\,{\rm pc}\,
(\theta_s/15^\prime)(D/3\,{\rm kpc})$ in our calculations.  The X-ray
emission from W44 is also elliptical in shape, although centrally
peaked, and lies entirely within the radio shell.  The elliptical
nature of the emission region implies that the models used, which are
spherically symmetric, cannot be completely valid. Nevertheless, as a
good first approximation, we compare the radially averaged surface
brightness profile from the \rosat\ PSPC with predictions from the models.

\subsection{The W44/PSR 1853+01 Association}
\par

In the PSR 1853+01 discovery paper (Wolszczan et al.\ 1991), the
arguments for associating the pulsar and the SNR W44 were first laid
out: positional coincidence, agreement in inferred distance, the youth
of the pulsar as indicated by the observed large-amplitude timing
noise, and agreement between the characteristic spin-down age of the
pulsar and the dynamical age of the remnant.  More recent research has
provided additional strong evidence to support this association. Frail
et al.\ (1996) have imaged the radio synchrotron nebula around PSR
1853+0.1, which they find to show an unusual cometary morphology with
the pulsar located near the narrow (southern) end of the nebula.  The
thermal pressure necessary to confine the radio nebula is roughly
$6\times 10^{-10}$ ergs cm$^{-3}$ which, while several orders of
magnitude larger than the pressure of the interstellar medium in
general, is within a factor of two of our pressure estimate for the
hot gas in W44 (\S 3.4 above).  This measurement can leave little
doubt that PSR 1853+0.1 lies within the hot X-ray emitting plasma of
W44 and that, consequently, the pulsar and SNR were formed in the same
supernova explosion.

\par

As we show below, {\it when} that SN explosion occurred is critical to our
understanding of the evolutionary state of W44.  One estimate is
provided by the spin-down age of the pulsar. The spin-down of pulsars
is believed to follow the relation $\dot\nu = - K \nu^n$, where $\nu$
is the rotation rate, $n$ is the braking index and $K$ depends on the
properties (such as the moment of inertia and magnetic field) of the
neutron star. A value of $n=3$ is expected if the pulsar's rotational
energy is lost purely through radiation from a dipole magnetic field.
Assuming $K$ and $n$ to be constant, one derives the age $t$ of the
pulsar
$$t = {P / \dot P \over (n-1)} \left[1-\left({P_0\over
P}\right)^{n-1}\right]$$ 
in terms of the initial spin period $P_0$ and
the current period $P$ and period derivative $\dot P$.  The braking
index has been measured for three young pulsars: the Crab (PSR
B0531+21), PSR B0540$-$69 and PSR B1509$-$58, and all show values less
than 3 for $n$: $2.51 \pm 0.01$ (Lyne, Pritchard, \& Smith 1993),
$2.20 \pm 0.02$ (Boyd et al.~1995), and $2.837 \pm 0.001$ (Kaspi et
al.~1994), respectively. Recently, Lyne et al.~(1996) measured the
braking index of the Vela pulsar which is roughly ten times older
than the pulsars mentioned above and in that sense most closely
resembles PSR 1853+0.1; they find a surprisingly low value for the
index $n=1.4\pm 0.2$. Nonetheless Lyne et al.~(1996) claim that the
age derived using this braking index and a low initial spin period
(20 ms, the estimated initial spin period of the Crab) results in a
value that is consistent with other estimates of the age of the Vela
SNR.

\par

The radio timing parameters of PSR 1853+0.1 are $P = 0.26743520599(6)$
s and $\dot P = (208.482 \pm 0.006) \times 10^{-15}$ s s$^{-1}$
(Wolszczan 1995).  Assuming a low initial spin period of 20 ms we
estimate an age for PSR 1853+0.1 of $2.65 \times 10^4$ yr with
$n=2.5$. If $n=1.5$, the age estimate is increased significantly to
$5.9 \times 10^4$ yr. In order for PSR 1853+0.1 to be younger than
$\sim$10,000 yr, then it must have been born as a slow rotator with a
spin period \ga 200 ms or have undergone an unusual spin-down
history. Although not inconceivable, this would make the pulsar in W44
different from the other known pulsars in SNRs discussed above.

\subsection{White \& Long Model}

\par

The WL similarity solution for the evolution of SNRs invokes a
multi-phase interstellar medium consisting of cool
dense clouds embedded in a tenuous intercloud medium.  The blast wave
from a SN explosion propagates rapidly through the intercloud medium,
in the process engulfing the clouds. In the model, these clouds are
destroyed by gradually evaporating on a timescale set by the saturated
conduction heating rate from the post-shock hot gas. Since this
timescale can be long, it is possible for cold clouds to survive until
they are well behind the blast wave which as they evaporate can
significantly enhance
the X-ray emission from near the center of the remnant.

\par

The timescale for cloud evaporation is one of the two parameters in
the WL model in addition to the three parameters which characterize
the standard Sedov solution (explosion energy $E_0$, ISM density
$n$, and SNR age $t$).  This timescale, which is expressed as a
ratio of the evaporation timescale to the SNR age, $\tau_e \equiv t_{\rm
evap}/t$, can depend on various factors, such as the composition of
the clumps and the temperature behind the shock front. The other new
parameter, $C$, represents the ratio of the mass in clouds to the mass
in intercloud material.  For appropriate choices of these two new
parameters, the model can produce a centrally peaked X-ray emission
morphology. Alternatively, other choices of the $\tau_e$ and $C$ can
reproduce the standard Sedov solution.  This model has been
applied to the centrally-peaked remnants W28 and 3C400.2 (Long et
al.\ 1991), \nocite{LONG91} as well as to CTA1 (Seward, Schmidt \&
Slane 1995). \nocite{SEWA95}

\par

We searched the $C$-$\tau_e$ plane of parameter space to determine which
values gave a good match to the W44 radial X-ray surface-brightness
profile. We integrated the differential equations for the WL
similarity solution to obtain the radial run of temperature and
density throughout the interior of the remnant.  These functions were
normalized to their values at the shock front. The temperature at the
shock front $T_s$ was related to the emission-measure-weighted
temperature $\langle T \rangle$ (which we measure) using equation (23)
in WL.  The density at the shock front $n_s$ was scaled to match the
observed emission measure of X-ray emitting gas in W44. For each set
of $C$ and $\tau_e$ values, appropriate values for $T_s$ and $n_s$ were
calculated.  With these values and the radial run of temperature and
density, it was possible to calculate the detailed radial X-ray
surface-brightness profile. For each radial bin in the SNR model, a RS
plasma model of appropriate temperature was calculated,
and the resulting photon spectrum was multiplied by the
energy-dependent ISM absorption function assuming our best-fit
column density. The absorbed spectrum was convolved with the PSPC
efficiency and spectral resolution functions and then projected to the
plane of the sky.  This was iterated over all radial bins of the model.

\par

Values of $C$ and $\tau_e$ in the ratio of approximately 2.5:1 for 
$5\: \la C\: \la 100$ provided reasonable profiles. In Figure~3 we show
the observed PSPC surface-brightness profile along with several
representative WL models. The dashed curves bracket the range of
acceptable solutions: the top one is too centrally peaked, while the
bottom one is too limb-brightened.  The three curves near the center
show examples of good fits.

\par

The dependence of remnant age $t$ on shock radius and temperature
$T_s$ in the WL model is identical to that of the Sedov solution:

$$ t = 5490\,{\rm yr}\,\biggl({\theta_s \over 15^\prime}\biggr)\,
  \biggl({D \over 3\,{\rm kpc}}\biggr)\,
  \biggl({kT_s \over 1\,{\rm keV}}\biggr)^{-1/2},$$

\noindent
which explicitly includes the functional dependence on distance $D$.
For the allowed range of $C$ and $\tau_e$ values, the shock
temperature varies between 0.53 keV and 0.95 keV, including the 
observational error on $kT$. This yields an age for W44 between 5600
yr and 7500 yr, which is similar to previous estimates of the age of
W44 based on application of the WL model (Rho et al.~1994).  The
square root dependence of $t$ on $kT_s$ means that the temperature
would have to be an order of magnitude less than the value we actually
measure to increase the remnant's age by a factor of 3. We can think
of no systematic effect in our data or analysis that could result in
such an enormous change in the mean temperature of W44.  Note that the
age of the remnant in this scenario also depends on distance. However,
in order for W44 to be $\sim$20,000 yr old, the remnant would need to
be 2.5 times further away than the accepted distance of 3 kpc. This
too is highly unlikely.

\par

Although this model appears to reproduce the intensity and morphology
of W44, it predicts an age that is much less than the characteristic
age of the associated pulsar.  In addition we also find that the
estimated initial explosion energies of the acceptable WL models is
rather small: $(0.11-0.16)\times 10^{51}$ ergs. These two results
considerably weaken the plausibility of the WL model as an accurate
description of the SNR W44, particularly in comparison to the model we
discuss next.

\subsection{Radiative Shock Model}

To study this alternative evolutionary scenario quantitatively we use
a one-dimensional, spherically symmetric, hydrodynamic shock code
(Hughes, Helfand, \& Kahn 1984). We include radiative cooling
parameterized by temperature as in Raymond, Cox, \& Smith (1976) for
material with solar abundances.  Models were generated for a range of
values for the initial explosion energy $E_0$ and ambient ISM Hydrogen
number density $n$, assumed to be homogeneous, isotropic, and free of
magnetic fields.  For completeness we have considered two extreme
cases for the exchange of energy between the shock-heated ions and
electrons: (1) rapid equilibration in which the electrons and ions
attain the same temperature instantaneously at the shock front and (2)
equilibration on a timescale set by Coulomb collisions. We find no
difference between these cases for the models that best describe W44
and so only quote results for the Coulomb equilibration models.

\par

The hydrodynamic calculation was initiated using a homologously
expanding, uniform density shell of ejecta with a total mass of 10
$M_\odot$ extending over a small spatial extent (from the center of
the explosion to a radius of 0.2 pc). At the ages of interest for our
modeling of W44, the reverse shock has passed completely through the
ejecta, fully thermalizing it.  It is well known that a decelerating
ejecta shell is Rayleigh-Taylor unstable and observations of young
ejecta-dominated SNRs such as Cas A and Tycho show clear evidence that
this instability indeed operates in nature. The effect of the
instability is to cause significant clumping of the ejecta shell that
ultimately results in its fragmentation and disruption. Our simple
one-dimensional calculation is unable to model this effect.  However,
our interest is studying the onset of dense shell formation at the
blast wave and not the fate of the SN ejecta, so this limitation of
our calculation is not important.  When calculating the projected
surface brightness we remove radial bins containing the modeled
ejecta, replacing them with an extrapolation of the temperature and
density profile from the solution further out, guided by the radial
run of temperature and density expected from the Sedov (1959)
solution.  This effectively removes the ejecta from the
calculation. In practice, much of the ejecta should remain within the
interior of the remnant and, by increasing the metallicity of the gas
there, enhance the central X-ray emission.  Whether the
centrally-peaked brightness of W44 and other SNRs in this class can be
explained, at least, partially by enhanced metallicity in the remnant
interior is beyond the scope of the current study. We will be
exploring this issue in future work by searching for abundance
gradients using spatially resolved X-ray spectral data.

\par

We initially explored values for $E_0$ and $n$, searching for model
remnants that attained radii between 12 pc and 14 pc in ages of from
19,000 yr to 25,000 yr.  This requirement largely constrained the
ratio of $E_0/n$ to be $\sim$$(0.2-0.4) \times 10^{51}$ ergs
cm$^{3}$. Next this range of model parameters was explored more finely
in order to find model SNRs that reproduced both the measured X-ray
intensity and a centrally-peaked surface-brightness profile. The model
temperature and density profiles were projected as above using the
best-fit ISM column density ($N_{\rm H} = 1.0\times 10^{22}$
atoms~cm$^{-2}$) to obtain surface-brightness profiles in terms of
PSPC counts s$^{-1}$ arcmin$^{-2}$ for comparison to the
data. Appropriate values of $E_0$ in the range $(0.1-2) \times
10^{51}$ ergs and $n$ in the range $0.25-11$ cm$^{-3}$ were
considered.  The best fit solutions to the W44 data were obtained for
$E_0 \approx ( 0.7 - 0.9 ) \times 10^{51}$ ergs and $n \approx ( 3 - 4
)$ cm$^{-3}$.  Figure~4 shows the radial X-ray surface-brightness
profiles of several of these acceptable models.

\par

The curves (labeled ``a'' and ``b'') show how the X-ray surface
brightness profile varies with age. The top curve is the model with
$E_0 = 0.9 \times 10^{51}$ ergs and $n = 3.0$ cm$^{-3}$ at 19,400 yr
and the bottom one is the same case at 25,200 yr.  In these models
there is a dense shell of radiatively cooled ISM at radii of 12.0 pc
(a) or 12.5 pc (b) at the outer edge of the remnant. Interior to this
the temperature rises slowly, increasing from about $10^6$ K just
inside the radiative shell to $10^7$ K near the center.  Over the same
radial range the density shows a gradient of the opposite sign,
decreasing from the edge of the remnant in toward the center.  The
centrally bright profile is a result of absorption by the large column
density to W44 of the soft X-rays from near the remnant edge. The
harder photons from the hotter gas in the interior are preferentially
less absorbed and thus, although the matter density is less there, it
has a higher observed emissivity.  The other curves in Figure~4 show
the X-ray brightness profiles of models with other values for $E_0$
and $n$.  (Note that we ran our shock model for the a fixed explosion
energy and age and found that different ISM densities yielded final
remnant radii that differed by the observed ellipticity of W44 but
still yielded roughly centrally-peaked morphologies.)

\par

We estimated nonequilibrium ionization effects on the broadband X-ray
surface brightness profile of the radiative phase model in the
following manner.  For each interior radial shell in the SNR model, we
integrated the time history of electron density to form the radial run
of ionization timescale, $\tau_i$. The values of $\tau_i$ so
determined plus the final state value of $kT$ were used with the
single-timescale, single-temperature NEI model to predict the X-ray
emissivity through the remnant interior and then the projected surface
brightness profile in the manner described previously.  (This
approximation differs from a full-up NEI model only to the extent that
the temperatures of individual radial shells may have changed with
time. We verified that this was a small effect by comparing the final
state temperature in radial bins of the model with the time-averaged
temperature and found that they differed only slightly, the
time-averaged temperature being somewhat higher.)  The NEI brightness
profile for one particular model is shown in Fig.~4. It differs from
the corresponding equilibrium ionization case by $<$10\%, confirming
that NEI effects on the modeled brightness profile are minor and that
the results presented above are robust. We note that the
emission-measure weighted mean ionization timescale of the NEI model
shown in Fig.~4 is $\sim$$9\times 10^{11}$ cm$^{-3}$ s, which is in
reasonable agreement with the observed value for W44 given the
simplicity of the model.

\par

It has occasionally been suggested that shock models would provide a
poor explanation of the centrally-peaked SNRs since such models are
expected to show significant radial temperature gradients. The
emission-measure-weighted projected temperature of our model labeled
``a'' in Figure~4 does show a larger variation than the 10\%--20\%
variation in temperature from the PSPC data (\S 3.2). The projected
temperature in this model varies by about $\sim$50\% from the center
to the edge.  A possible solution to this problem could be provided by
including thermal conduction processes (e.g., Cui \& Cox 1992; Shelton
\& Cox 1995; Smith 1996) which can have the effect of smoothing out
strong temperature gradients.  Although promising, studies of this
kind are beyond the scope of the current article and are deferred to
future research.

\par

The emission-measure-weighted average temperatures of the acceptable
shock models, which are in the range 0.4 keV to 0.5 keV, are also
slightly lower than the observed average temperature value of $kT=0.88
\pm 0.12$ keV from the current data. An independent estimate of the
mean temperature for the hot plasma in W44 was obtained by Harrus et
al.~(1996) from a preliminary analysis of \asca\ data, which gave a
somewhat lower result $kT = 0.5 \pm 0.2$ keV. Averaging these two
independent measurements produces a result for the temperature of W44
that is slightly more consistent with the model predictions, although
still somewhat high. Clearly more careful analysis of the \asca\
data, plus additional modeling along the lines mentioned in the
preceding paragraph, will be crucial in furthering our understanding
of the nature of W44 and the other filled-center remnants.

\par

All things being considered it is remarkable that this simple model, a
function of only three parameters, reproduces both the observed
intensity of the remnant and, at least to first approximation, the
centrally peaked X-ray brightness profile. The derived parameters, in
particular an explosion energy of $\sim$$0.9 \times 10^{51}$ ergs, are
physically quite plausible. And since the remnant's age is similar to
the characteristic age of the pulsar, there is no need to invoke an
unusual evolutionary scenario for the spin-down of the pulsar.
Because of these considerations, we favor this, the radiative phase
shock model, for the interpretation of the center-filled X-ray
emission from W44.

\par

This model also accounts quite naturally for the massive,
high-velocity shell of H {\sc i} gas that has been observed to
surround the X-ray emitting region of W44 (Koo \& Heiles 1995).  The
shell's velocity is estimated to be 150 km s$^{-1}$ and its mass
$\sim$350 $M_\odot$ with large uncertainties. Although Koo
\& Heiles did explore the possibility that the H {\sc i} shell was a
consequence of W44 being in the radiative phase of evolution, they
rejected this interpretation because of a belief that the
centrally-peaked X-ray emission from W44 could not be reconciled with
the standard radiative model.  Our work here has shown that this is
just not true; in fact the radiative phase model is the preferred
explanation for the nature of W44. The cool shell in our radiative
phase models has a mass within the range of $\sim$550--900 $M_\odot$
and expansion velocities of 100--120 km s$^{-1}$ (depending on $E_0$ and
$n$), values that are in good agreement with the H {\sc i}
observations.

\section{Summary}

In this article we have presented an analysis of X-ray data from the
\einstein\ SSS, the \rosat\ PSPC, and the \ginga\ LAC on the
supernova remnant W44.  These spectral data are well described by a
single-temperature, single-timescale nonequilibrium ionization model
with temperature $0.88\pm 0.014$ keV and ionization timescale
$(2.0^{+4.3}_{-0.7})\times10^{11}$ cm$^{-3}$~s, observed through a
large absorbing column density: $N_{\rm H} = (1.0^{+0.6}_{-0.2})\times
10^{22}$ atoms~cm$^{-2}$. All elemental abundances are close to the 
solar values, with iron showing possibly significant depletion.

\par

Morphologically, W44 belongs to the class of SNRs that have clear
shell-like structures in the radio but are centrally-peaked in the
X-ray band and exhibit thermal X-ray spectra.  We have examined in detail two
proposed scenarios for the origin of this structure: (1) a model
specifically developed for application to this class of remnants
invoking a long evaporative timescale for the destruction of clouds
engulfed by the SN blast wave (White \& Long 1991) and (2) a model of
remnant evolution in a homogeneous medium during the post-Sedov phase
of development when radiative cooling at the shock front has become
important.  Because W44 is such a well-studied object there is a
wealth of information available on it.  The distance is accurately
known and, since there is an associated pulsar, we have an independent
estimate of the age of the remnant.  Our measurement of the mean
temperature of the hot plasma in W44 from the X-ray observations,
coupled with its age and size, is what provides the strongest
constraints on the evolutionary state of the remnant.

\par

Taking the size and temperature as the fundamental observables, we
find that the WL model predicts an age of 5600--7500 yr, which is
incompatible with the characteristic age ($\sim$20,000 yr) of the
associated pulsar PSR 1853+0.1.  This is considerably greater than any
discrepancy that could be resolved though errors in the distance to
W44 or the X-ray temperature measurements.  It would require that the
pulsar in W44 to have been born as a slow rotator ($P_0$ $\ga$ 200 ms)
or to have undergone an unusual spin-down history. The WL model also
predicts an unusually low explosion energy $\la$$0.2\times 10^{51}$
ergs for the core-collapse SN that is believed to have formed W44 and
PSR 1853+0.1. These two observationally-derived conclusions are the
basis on which we reject this model as the explanation of this
center-filled remnant in favor of the radiative shock phase model.
However, we also wish to highlight the astrophysical implausiblity of
a major assumption of the WL model: i.e., that the ISM clouds engulfed
by a SNR should survive being crushed by the blast wave and linger
within the interior to be gently evaporated away on timescales that
are many times the age of the remnant. We do not reject the WL model
because we consider a cloudy ISM unlikely, rather it is the {\it
timescale} for the destruction of those clouds that is at issue.

\par

Our alternative scenario for W44 has the remnant in the post-Sedov
radiative phase of evolution. In this case we find that a
centrally-peaked morphology and nearly uniform temperature profile can
occur for models that are roughly 19,000 yr to 25,000 yr old for
reasonable explosion energies ($\sim$$0.9\times 10^{51}$ ergs) and ISM
densities of 3--4 cm$^{-3}$ (assumed uniform and
homogeneous). These are not outrageously large values for the ambient
density; rather they are very similar to the ambient densities
estimated around the SNRs N132D and N49 in the Large Magellanic Cloud
(Hughes 1987; Vancura \etal\ 1992).  The reason for these largish
densities is attributed to the presence of nearby molecular clouds for
the LMC SNRs (Banas et al.~1997).  A similar explanation is likely for
W44, since it too appears to be associated with molecular emission
(Wootten 1977).  Our model accounts naturally for the high velocity
shell of H {\sc i} gas which surrounds the X-ray emitting gas in W44.
Finally, the radiative phase scenario is qualitatively consistent with
the gross characteristics of the VLA radio image: filamentary radio
emission concentrated at the rim rather than the remnant interior.
 
\par

Additional research on W44 should be directed toward obtaining
spatially resolved measurements of temperature and elemental
abundance. This will be a challenging measurement to make since both
the radiative phase model and the current data support the presence of
only a modest radial variation in temperature. We are pursuing this
issue further with the available \asca\ data.  It is also interesting
to note that a more accurate estimate of the remnant's age may become
available in the future.  Recently, Frail et al.\ (1996) have argued
that the pulsar in W44 is moving toward the south with a speed of
approximately 375 km s$^{-1}$. Since the likely location of the SN
explosion which gave rise to the pulsar is known (i.e., the centers of
the radio continuum, X-ray, and H {\sc i} images of W44), measurement of the
proper motion of PSR 1853+0.1 (estimated to be of order 25 mas
yr$^{-1}$) would provide a distance-independent determination of the
age of the pulsar and the SNR.  As we have shown in this article, such
a definitive measurement would yield a crucial constraint on models
for the evolutionary state of W44 and should be pursued.

\par
\vskip 24 pt

We thank Fred Seward, Pat Slane, Olaf Vancura, and David Helfand for
useful discussions and comments during the course of this project.
Our research made use of data obtained from the High Energy
Astrophysics Science Archive Research Center Online Service, provided
by the NASA/Goddard Space Flight Center.  K.\ P.\ S.\ acknowledges the
hospitality of the High Energy Astrophysics Division of the Center for
Astrophysics and thanks the Smithsonian Institution for funding his
visit to the CfA.  This research was supported in part by NASA under
grants NAG8-670, NAG8-181, and NAG8-287 and by Smithsonian Institution
funds from the International Exchange Program and the Predoctoral
Program through a Fellowship awarded to I.~M.~H.

\clearpage

\centerline{\bf Table 1. Elemental Abundances of W44}
\vspace{0.2cm}
\centerline{\begin{tabular}{lccccc} \hline\hline \\[-3mm]
        & Best fit            & Random 
  & \multicolumn{3}{c}{Error due to variation with:} \\ \cline{4-6}
Species & relative to $\odot$ & Error & \quad $N_{\rm H}$ & $n_e t$ 
   & $kT$  \\ [1mm] \hline \\[-3mm]
Ne & 0.76 & $^{+0.35}_{-0.35}$ & $^{+0.18}_{-0.78}$
  & $^{+0.20}_{-0.15}$ & $^{+0.40}_{-0.10}$ \\[1mm]  
Mg & 0.90 & $^{+0.33}_{-0.25}$ & $^{+0.80}_{-0.10}$ 
  & $^{+0.50}_{-0.10}$ & $^{+0.60}_{-0.10}$ \\[1mm]
Si & 0.70 & $^{+0.21}_{-0.16}$ & $^{+0.25}_{-0.05}$ 
  & $^{+0.15}_{-0.05}$ & $^{+0.20}_{-0.10}$\\[1mm]
S  & 0.64 & $^{+0.28}_{-0.20}$ & $^{+0.15}_{-0.05}$ 
  & $^{+0.15}_{-0.05}$ & $^{+0.15}_{-0.05}$  \\[1mm]
Ar & 0.24 & $^{+0.40}_{-0.15}$ & $^{+0.15}_{-0.05}$ 
  & $^{+0.10}_{-0.05}$ & $^{+0.15}_{-0.05}$\\[1mm]
Fe & 0.07 & $^{+0.08}_{-0.07}$ & $^{+1.10}_{-0.07}$ 
  & $^{+0.35}_{-0.07}$ & $^{+0.65}_{-0.07}$ \\[1mm] \hline
\end{tabular}}

\clearpage

\begin{figure}
\caption{
The top panel shows the X-ray spectra of W44 from
the {\it ROSAT} PSPC, {\it Einstein} SSS, and {\it Ginga} LAC with
the best-fit single-temperature, single-timescale nonequilibrium
ionization model. The residual spectrum (observed data minus best-fit
model) is shown in the bottom panel.}
\end{figure}

\begin{figure}
\caption{
The effect of variation in column density
($N_{\rm H}$), temperature ($kT$), and ionization timescale ($n_e t$)
on the best-fit emission measures of the various species considered
from fits to the W44 X-ray spectra using the nonequilibrium ionization
model. The top three panels show the change in $\chi^2$ with the
independent variation of each spectral parameter.  The 90\% confidence
interval (for three interesting parameters) is shown in each panel as
a dashed line ($\chi^2_{min} + 6.25$).  The variations of elemental
abundance (relative to the appropriate solar value) are shown in the
bottom panels. The error bars are plotted at the best-fit values and
represent the random error at 90\% confidence.}
\end{figure}

\begin{figure}
\caption{
Radial X-ray surface brightness profile of W44 from the
{\it ROSAT} PSPC (shown as the 7 data points with statistical error
bars) compared to the White and Long (1991) similarity solution for
the evolution of SNRs in a cloudy interstellar medium. The solid
curves show acceptable fits while the dashed curves indicate the
extremes of the allowed solutions.  These models all predict a young
age (5600--7500 yr) and low explosion energy (0.11--0.16) $\times$
$10^{51}$ ergs for W44.}
\end{figure}

\begin{figure}
\caption{
Radial X-ray surface brightness profile of W44 from the
{\it ROSAT} PSPC compared to radiative-phase shock models. The curves
are labeled with the explosion energy (in units of $10^{51}$ ergs)
and ambient ISM density (cm$^{-3}$) of the model as ($E_0$, $n$). The
two curves delineating the dotted region labeled ``a'' and ``b'' show
how the profile varies with remnant age from 19,000 yr (a) to 
25,000 yr (b). All these models are consistent with the spin-down age of
the associated pulsar PSR 1853+01, $P/2\dot P \sim 20000$ yr.}
\end{figure}

\clearpage
\begin{figure}
\plotone{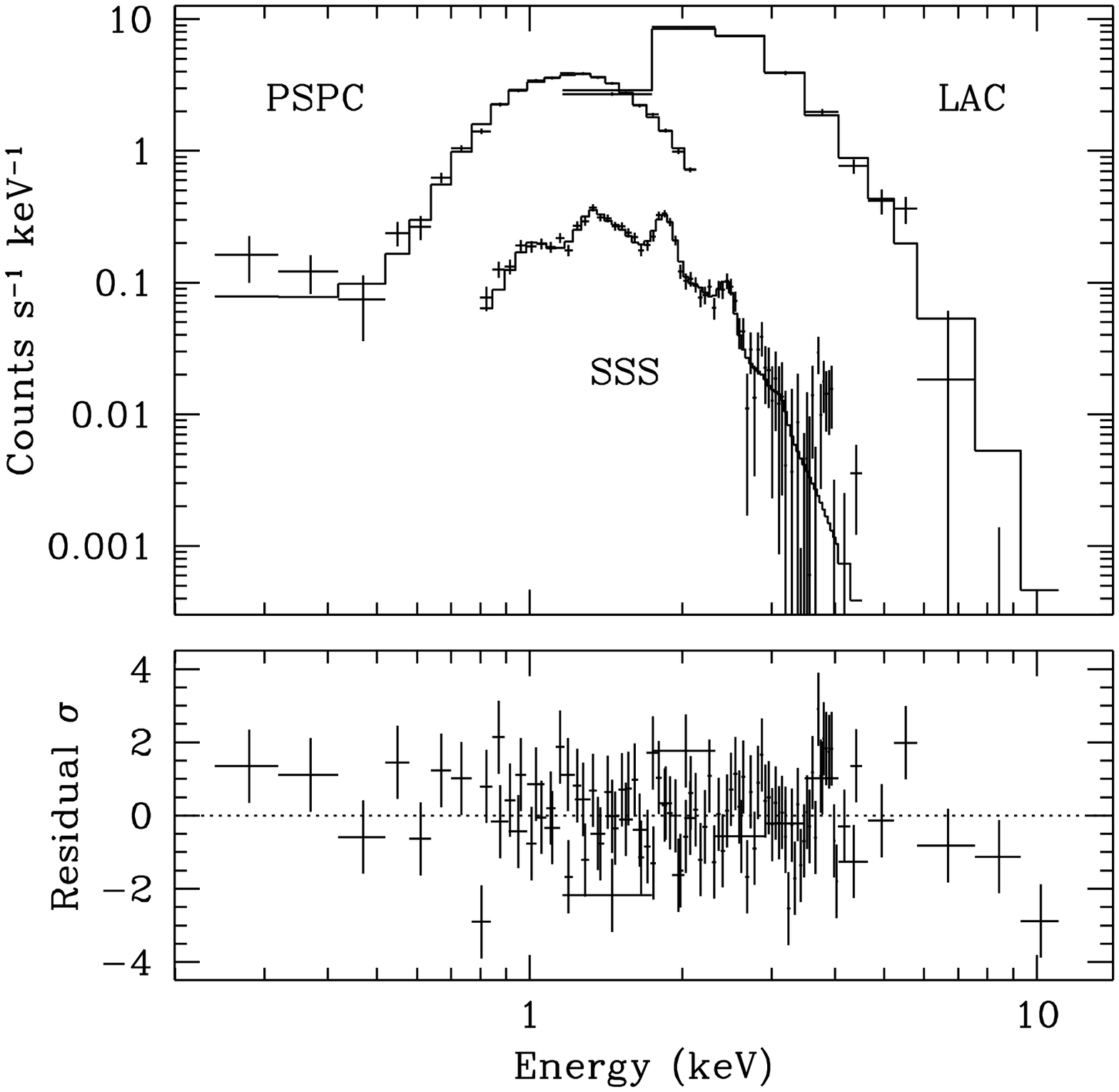}
\end{figure}

\clearpage
\begin{figure}
\plotone{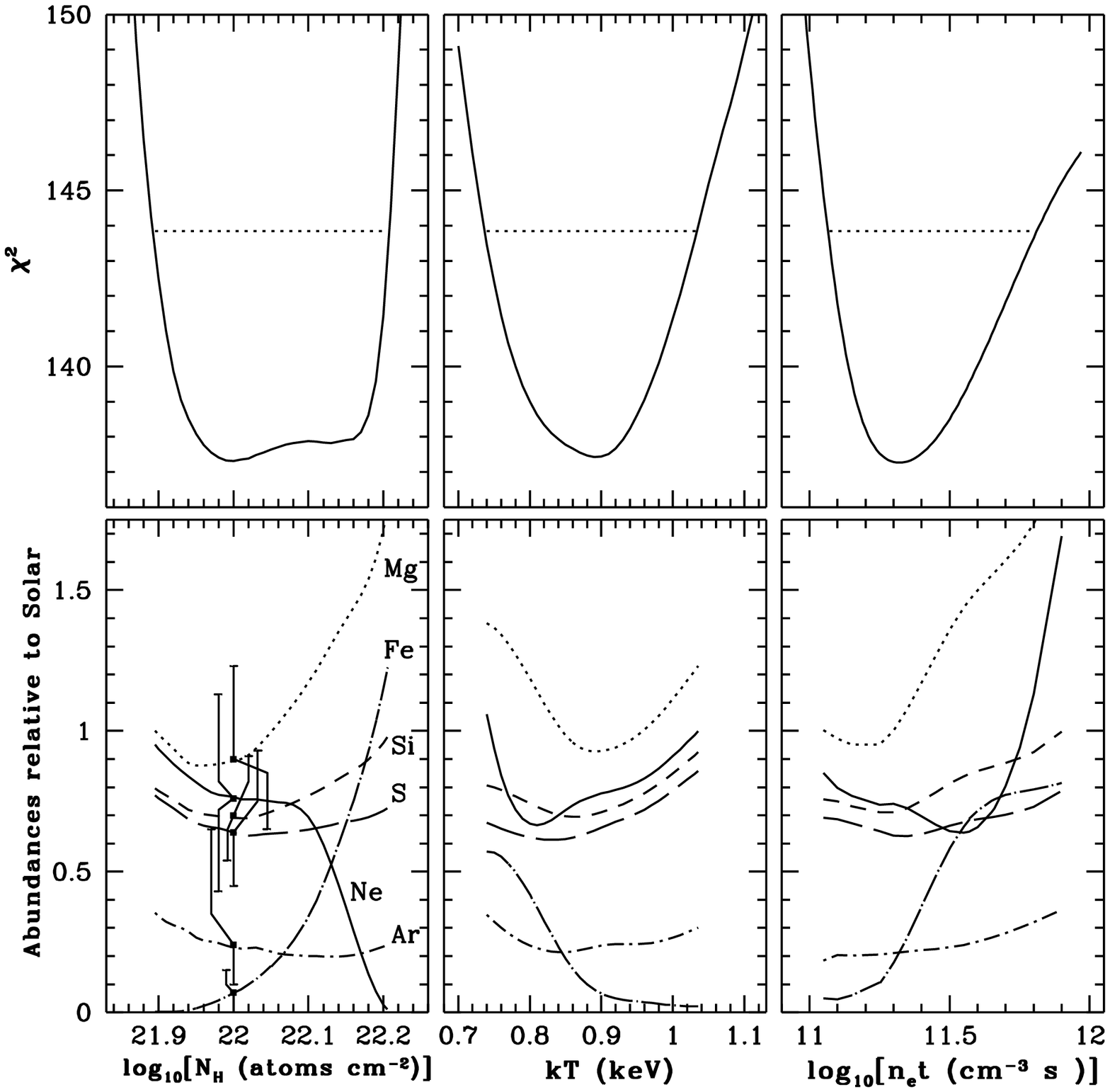}
\end{figure}

\clearpage
\begin{figure}
\plotone{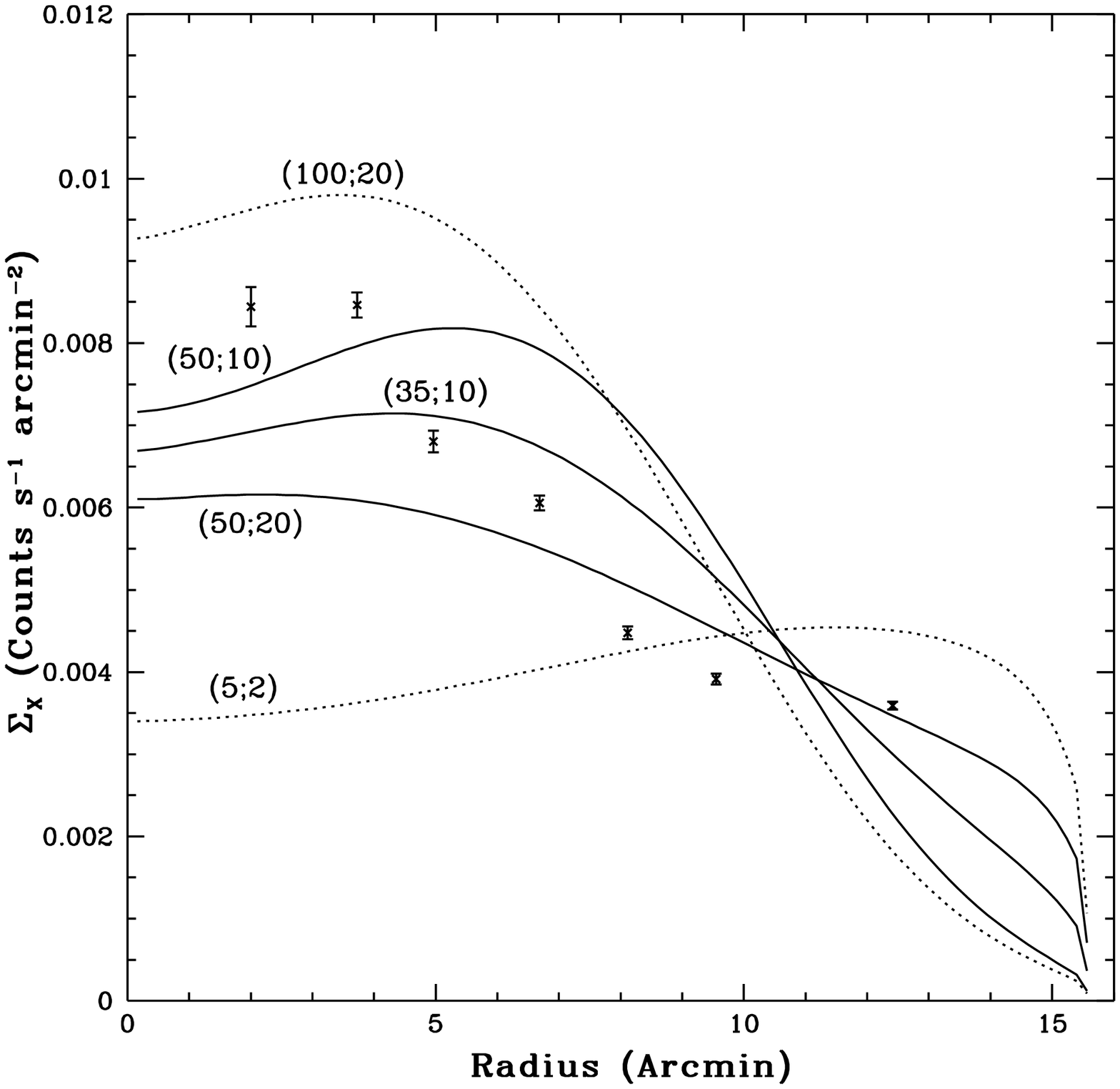}
\end{figure}

\clearpage
\begin{figure}
\plotone{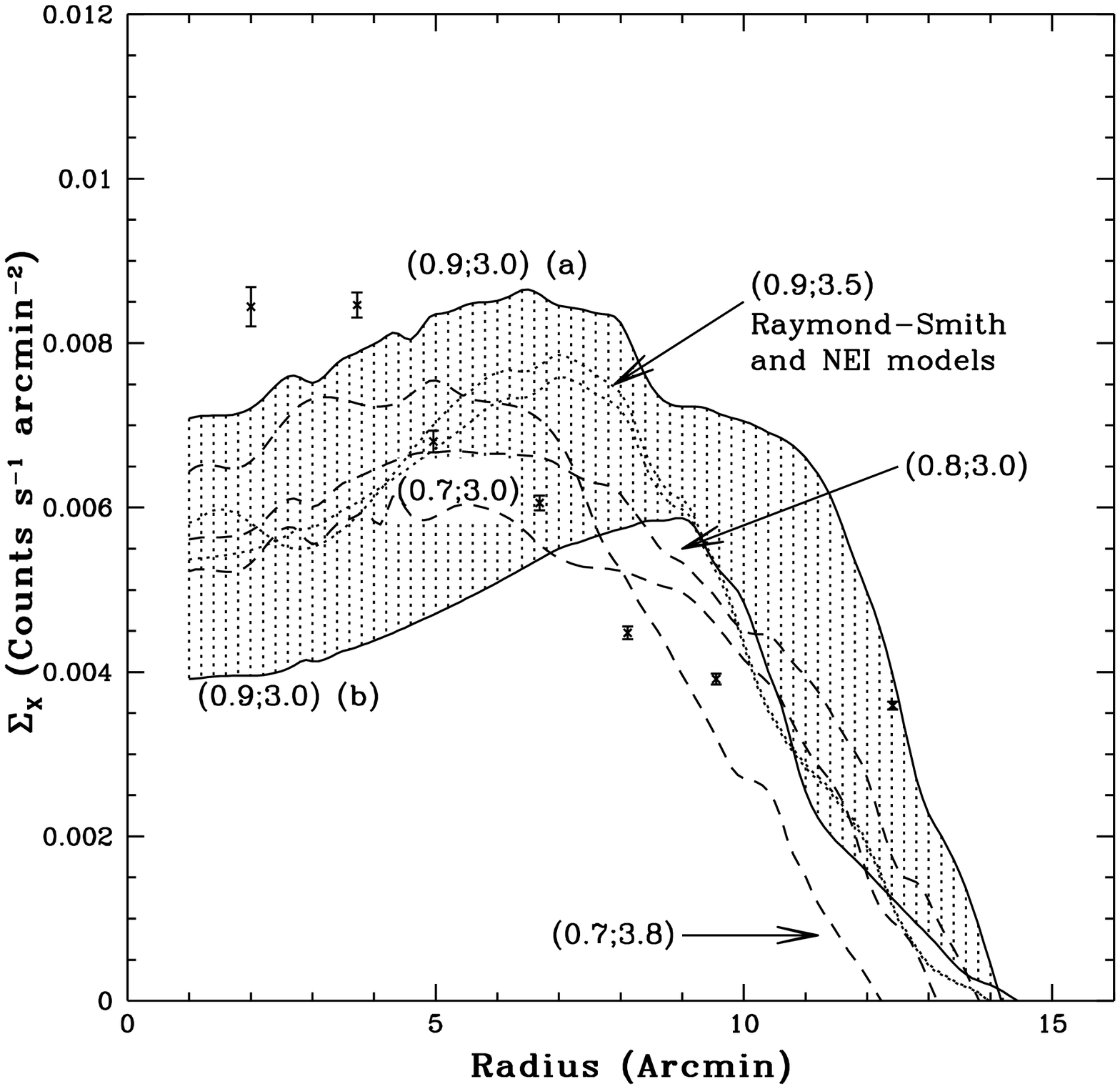}
\end{figure}

\end{document}